\documentclass[onecolumn,
               preprintnumbers,
               prl]{revtex4}
\usepackage{graphicx, subfigure}
\usepackage{amssymb, amsmath,amssymb,amsfonts}
\usepackage{dcolumn}
\usepackage{bm}
\usepackage{color}
\usepackage[utf8]{inputenc}

\newtheorem{theorem}{Theorem}

\usepackage[framemethod=tikz]{mdframed}  %in order to highlight changes: 
\begin{document}

\title{The theory of Turing patterns on time varying networks}

\author{Julien Petit$^{1,2}$, Ben Lauwens$^2$, Duccio Fanelli$^{3,4}$, {Timoteo Carletti}$^1$}
\affiliation{$^1$naXys, Namur Institute for Complex Systems, University of Namur, Belgium \\
$^2$Department of Mathematics, Royal Military Academy, Belgium\\
$^3$Dipartimento di Fisica e Astronomia and CSDC, Universit\`a degli Studi di Firenze, Italy\\
$^4$INFN Sezione di Firenze, Italy
}

\begin{abstract}
The process of pattern formation for a multi-species model anchored on a time varying network is studied. A non homogeneous perturbation superposed to an homogeneous stable fixed point can amplify, as follows a novel mechanism of instability, reminiscent of the Turing type, instigated by the network dynamics. By properly tuning the frequency of the imposed network evolution, one can make the examined system behave as its averaged counterpart, over a finite time window. This is the key observation to derive a closed analytical prediction for the onset of the instability in the time dependent framework. Continuously and piecewise constant periodic time varying networks will be analysed, to set the ground for the proposed approach. The extension to non periodic settings will also be discussed.
\end{abstract}

%\pacs{89.75.Hc, 89.75.Kd, 89.75.Fb}

\maketitle

%\section{Introduction}
%\label{sec:intro}

Spatially extended systems can spontaneously yield a multitude of patterns, archetypal imprint of an inherent drive to coordinated self-organisation~\cite{pikovsky,murray2,cross}. In many cases of interest, the subtle  interplay between nonlinearities and diffusion seeds a symmetry breaking instability (discovered by Turing in his pioneering work on morphogenesis),  which paves the way to the emergence of a rich and colourful gallery of patchy motifs~\cite{turing,mimura, baurmann}.  Distinct populations of homologous constituents often interact via an intricate architecture of nested couplings, which can be adequately represented as complex heterogeneous networks. Elaborating on the mechanisms that instigate pattern formation for reaction-diffusion systems hosted on discrete network-like supports is hence central to a wide gallery of phenomena of broad applied and fundamental impact~\cite{nakao}. 

Topology is known to play, in this respect, a role of paramount importance. Surprisingly, patterns can rise on a directed support~\cite{directed}, even if they are formally impeded on a regular, continuum or discrete, spatial medium.  Further, self-organisation may proceed across multiple, interlinked networks, by exploiting the multifaceted nature of resources and organisational skills. Multiplex networks in layers have been therefore introduced as a necessary leap forward in the modelling effort: these are particularly relevant to transportation systems, the learning organisation in the brain and to understanding the emergent dynamics in social communities. Interestingly, the interaction between adjacent layers can yield self-organised patterns which are instead impeded in the limit of decoupled layers~\cite{AsllaniPRE2014}. Patterns on individual layers can also fade away due to cross-talking between contiguous slabs. All these examples, point to the pivotal role exerted by the topology of the underlying network in shaping the system response and driving the onset of non-trivial instabilities. 

In several realms of application, the networks that specify the routing of the spatial or physical interactions are not static, but, instead, do evolve in time~\cite{holme2012,holme,delvenne,Masuda}. This is an important additional ingredient that should be eventually accommodated for, so to improve on our current understanding of the mechanisms that are ultimately responsible for the appearance of structured patterns across heterogenous networks. In the framework of temporal networks, diffusion processes~\cite{perra} as well as random walks~\cite{starnini2012,MasudaKlemmEguiluz2013} have been studied in recent years, mostly under the assumption of piecewise constant time varying networks \cite{Vankeerberghen2014,Liberzon1999,Stilwell2006}: the topology of the assigned connections is fixed over a finite window in time, whose duration can either depend on the dynamics of the system or exogenously controlled. The couplings are then instantaneously modified (created/destroyed/rewired) and the ensuing network frozen for the subsequent time interval. 
Synchronisation phenomena have also been considered for nonlinear oscillators~\cite{Barreto} hosted on time varying networks, as well as for systems displaying  generic
reactions terms~\cite{Kohar,Stilwell2006,Boccaletti2006}. 

In this Letter we consider for the first time the process of pattern formation for reaction-diffusion systems anchored on networks that evolve over time. To this end, we will initially inspect the case of a network that is periodically rearranged in time. We will in particular  prove that a symmetry breaking instability which anticipates the pattern derive can be incited, by properly tuning the frequency of the imposed network dynamics. 
The proposed framework includes the case of a continuously time varying network  (links weights change as a smooth function of time) and the one where sudden switches between distinct discrete networks configurations are to be accounted for (links are suddenly created/destroyed/rewired). Surprisingly enough, patterns can emerge for a reaction-diffusion system hosted on a piecewise constant time varying network, also when they are formally impeded on each network snapshot. The extension to non periodic settings will be also analysed, allowing us to draw a general interpretative scenario. 

Imagine two different species living on a network that evolves over time and denote by $u_i$ and $v_i$ their respective concentrations, as seen on node $i$. Links can change their (positive) weights, they can fade away (the corresponding weight goes to $0$), or be created (a null weight is turned positive) or even rewired (a combination of both preceding moves). For the sake of simplicity, and without losing generality, we deal here with networks made of a constant number of nodes, that totals in $N$. The network structure is stored in a time varying $N\times N$ weighted adjacency matrix, whose elements are labelled $A_{ij}(t)$. The (weighted) connectivity, $s_i(t)$, of node $i$ is computed as  $s_i(t)=\sum_j A_{ij}(t)$. Species can relocate in space, as follows a standard diffusive mechanism, ruled by the (time dependent) Laplacian operator $L_{ij}(t)=A_{ij}(t)-s_i(t)\delta_{ij}$.  Remark that, for all $t$ and all $i$, $\sum_j L_{ij}(t)=0$, namely $\mathbf{L}(t)$ is a row stochastic matrix. When species happen to share the same node, they mutually interact via non linear terms that reflect the specificity of the problem at hand. 
We will deal at first with networks that are periodically reset in time, and show how adjusting the period of the modulation can stimulate the onset of the patterns.

The model can be mathematically cast in the form:   
\begin{eqnarray}
\dot{u}_i(t)&=& f(u_i,v_i) + D_u\sum_{j=1}^{N}{L}_{ij}(t/\epsilon) u_j(t) \nonumber\\
\dot{v}_i(t)&=& g(u_i,v_i) + D_v\sum_{j=1}^{N}L_{ij}(t/\epsilon) v_j(t)
\label{eq:reac_diffacc}
\end{eqnarray}
where $f$ and $g$ stand for the generic nonlinear reactions terms and $D_u$ and $D_v$ label the diffusion coefficients of species $u$ and $v$, respectively. The parameter $\epsilon$ can be tuned at will, so as to modify the 
period of the Laplacian oscillations in time. We will in particular denote with $T$ the period obtained for $\epsilon=1$. As we shall prove, by forcing the network to oscillate at high frequencies, one can prompt a 
Turing-like instability.  When the oscillations materialise as successive swaps between two (or more) network configurations, it is the frequency of the blinking that sets the patterns derive, also when Turing 
motifs cannot manifest for the model constrained on any of the considered (static) network snapshots.  As a further condition, preliminary to the forthcoming development, we will assume that system (\ref{eq:reac_diffacc}) admits an homogeneous stable fixed point. In other words, we shall require that $(u_i,v_i)=(\bar{u},\bar{v})$ exists  such that $\dot{u}_i=\dot{v}_i=0$, for all $i=1,\dots,N$. 

Define the averaged Laplacian $\langle\mathbf{L}\rangle=\frac{1}{T}\int_0^T \mathbf{L}(t) dt$
and introduce the averaged reaction-diffusion system:
\begin{eqnarray}
\dot{u}_i(t)&=& f(u_i,v_i) + D_u\sum_{j=1}^{N}\langle L_{ij}\rangle u_j \nonumber\\
\dot{v}_i(t)&=& g(u_i,v_i) + D_v\sum_{j=1}^{N}\langle L_{ij}\rangle v_j \, ,
\label{eq:reac_diffave}
\end{eqnarray}
 
A conventional linear stability analysis, calibrated so as to sense the spectral characteristics of the averaged network, can be performed and allows one to conclude on the possibility for 
patterns {\em \`{a} la Turing} in the framework of the time independent picture (\ref{eq:reac_diffave}). Imagine that conditions are met so to enable a symmetry breaking instability of the homogeneous stable fixed point, 
as triggered by an external non homogeneous perturbation. Then, it can be rigorously proven that $\epsilon^*>0$ exists such that Eqs.~\eqref{eq:reac_diffacc} exhibit patterns of the Turing type, for  $0<\epsilon<\epsilon^*$. Here,
$1/\epsilon^*$ acts as an effective high frequency drive for self-organised spatial motifs to develop. The idea of the proof is sketched in the following, further details are relegated in the Supplementary Information (SI).

Label $\tau= t/\epsilon$ and introduce the compact notation $\vec{x}=(u_1,\dots,u_N,v_1,\dots,v_N)$. Then, by virtue of the theorem of averaging one can show that $\epsilon^*>0$ exists such that for all $0<\epsilon <\epsilon^*$ and $\tau=\mathcal{O}(1/\epsilon)$: 
\begin{equation*}
\vec{x}(\tau)-\vec{y}(\tau)=\mathcal{O}(\epsilon)\, ,
\end{equation*}
where $\vec{y}(\tau)$ is the solution of the averaged system:
\begin{equation}
\label{eq:reac_diffaccy}
\vec{y}\,^{\prime}(\tau)=\epsilon\left[F(\vec{y}) +\langle\mathcal{L}\rangle \vec{y}\right]\, ,\quad \vec{y}(0)=\vec{x}_0\in \mathbb{R}^{2N}\, ,
\end{equation}
subject to the initial conditions $\vec{x}~(0)~=~\vec{y}~(0)$. Since the reaction term, $F(\vec{x})=(f(u_1,v_1),\dots,f(u_N,v_N),g(u_1,v_1),\dots,g(u_N,v_N))$, is time independent, the average solely affects the diffusive part, organised in the block matrix
%\begin{equation}
$\mathcal{L}(t) :=\left(
\begin{smallmatrix}
 D_u \mathbf{L}(t) & 0 \\ 0 &  D_v \mathbf{L}(t) 
 \end{smallmatrix}
\right)$.
%\label{eq:DLx}
%\end{equation}

The solutions of, respectively, the time dependent system (with $\epsilon \in (0, \epsilon^*]$)  and the averaged one, stay close for times $\mathcal{O}(1)$. Hence,
if an injected perturbation destabilises the homogenous solution of~\eqref{eq:reac_diffave}, the same holds when the perturbation is applied to system~\eqref{eq:reac_diffacc}. 
Remark however that the two systems behave similarly, only for a finite time span.  We are in fact unable to prove an {\em effective or exponential stability} \`{a} la Nekhoroshev. Stated differently, we cannot guarantee that 
 the smaller $\epsilon$ the longer the two systems agree. This in turn implies that the asymptotic patterns, which follows the initial instability and as obtained for respectively the averaged and the time dependent models, may, 
 in principle, differ. We will return later on characterising the critical threshold $\epsilon^*$. 
 
%\textbf{Periodic twin networks.}
%\section{An application: the periodic twin networks}
%\label{sec:appl}
We shall here discuss a specific application (termed the {\it twin network} case) aimed at clarifying the conclusion reached above. To this end we will consider the Brusselator model, a widely studied model of chemical oscillators. 
This choice amounts to setting  $f(u,v)=1-(b+1)u+cu^2v$ and $g(u,v)=bu-cu^2v$ where $b$, $c$ stand as free parameters. When the model is made spatially extended, the coaction of diffusion and reaction terms 
can disrupt the homogeneous fixed point, paving the way to the subsequent patterns derive. This occurs when the parameters are set so as to make the dispersion relation ($\max\Re \lambda$, the real part of the complex exponential growth rate $\lambda$) positive over a finite window of wave-numbers. When the diffusion takes place on a heterogeneous network, the dispersion relation is defined on a discrete support that coincides with the eigenvalues of the Laplacian operator. If the spectrum of the discrete Laplacian operator falls in a region where the continuous dispersion relation is negative, no instability can take place on the network, even if it can develop on a continuum support. This observation will be central for what follows.

\begin{figure*}[ht!]
\begin{center}
\includegraphics[width=0.6\textwidth]{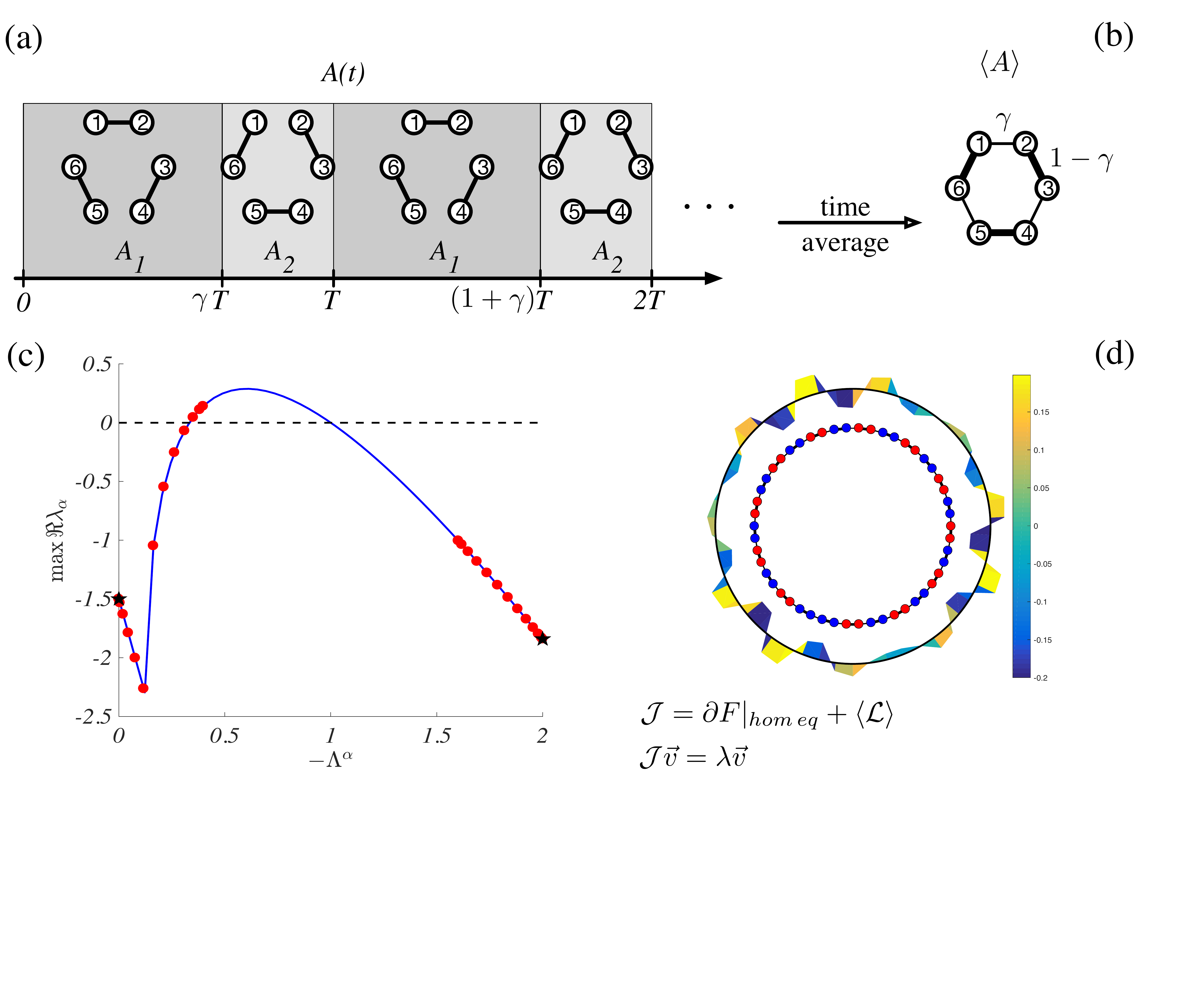}
\end{center}
\vspace{-2cm}\caption{Twin network. Panel a): $T$-periodic network built from two static networks, of adjacency matrices $\mathbf{A}_1$ and $\mathbf{A}_2$. Each network in this illustrative example
is made of  $N=6$ nodes. In the network stored in matrix $\mathbf{A}_1$, symmetric edges are drawn between the pairs $(1,2)$, $(3,4)$ and $(5,6)$. The second network,  embodied in matrix $\mathbf{A}_2$, links nodes $(6,1)$, $(2,3)$ and $(4,5)$.
For $t\in[0,\gamma T)$ the $T$-periodic network coincides with $\mathbf{A}_1$, $\mathbf{A}(t)=\mathbf{A}_1$, while in $[\gamma T,T)$ we set $\mathbf{A}(t)=\mathbf{A}_2$. The time varying network is then obtained by iterating the process in time. 
Panel b): the ensuing time average network $\langle \mathbf{A}\rangle =  \gamma \mathbf{A}_1+(1-\gamma)\mathbf{A}_2$. Panel c):  dispersion relation ($\max\Re \lambda_{\alpha}$ vs. $-\Lambda^{(\alpha)}$) for the average network (red circles), for each static twin network (black stars) and for the continuous support case (blue curve). Here, the  networks are generated as discussed above but now $N=50$. Panel d): patterns in the average network. Nodes are blue if they present an excess of concentration with respect to the homogeneous equilibrium solution ($(u_i(\infty)-\bar{u})\geq 0.1 $) and red otherwise ($(u_i(\infty)-\bar{u})\leq -0.1$). The outer drawing represents the entries of $\vec{v}$, the eigenvector of the Jacobian matrix $\mathcal{J}$ associated to the eigenvalues that yields the largest value of the dispersion relation.
The black ring stands for the zeroth level; red-yellow colours are associated to positive entries of $\vec{v}$, while blue-light blue refer to negative values. The reaction model is the Brusselator with $b=8$, $c=10$, $D_u = 3$ and $D_v = 10$.  The homogeneous equilibrium is $\bar{u}=1$ and $\bar{v}=0.8$. The remaining parameters are set to $\gamma=0.3$, $T=1$, $D_u = 3$ and $D_v = 10$.}
\label{fig:Fig1}
\end{figure*}

Let us thus consider two simple networks, made of $N$ nodes arranged on a periodic ring, and label with $\mathbf{A}_1$ and $\mathbf{A}_2$ their respective adjacency matrices. The $N$ 
nodes are connected in couples (so we will assume for simplicity $N$ even), via symmetric links. In the first case (the network that is encoded in $\mathbf{A}_1$)  
nodes $2k-1$ and $2k$, for $k=1,\dots,N/2$ are tight together. In the other, the couples are formed by nodes $(2k,2k+1)$ for $k=1,\dots,N/2-1$, with the addition of pair $(N,1)$ (see Fig.~\ref{fig:Fig1} panel a). 
Both networks yield an identical Laplacian spectrum: two degenerate eigenvalues are found,  $\Lambda^1=0$, with multiplicity $N/2$ (i.e. the number of connected components the network is made of) and $\Lambda^N=-2$, with multiplicity $N/2$. The parameters are set so that patterns cannot emerge for the Bussellator model defined on each of the networks introduced above. This is exemplified in Fig.~\ref{fig:Fig1} panel c.

\begin{figure*}[ht!]
\begin{center}
\includegraphics[width=0.7\textwidth]{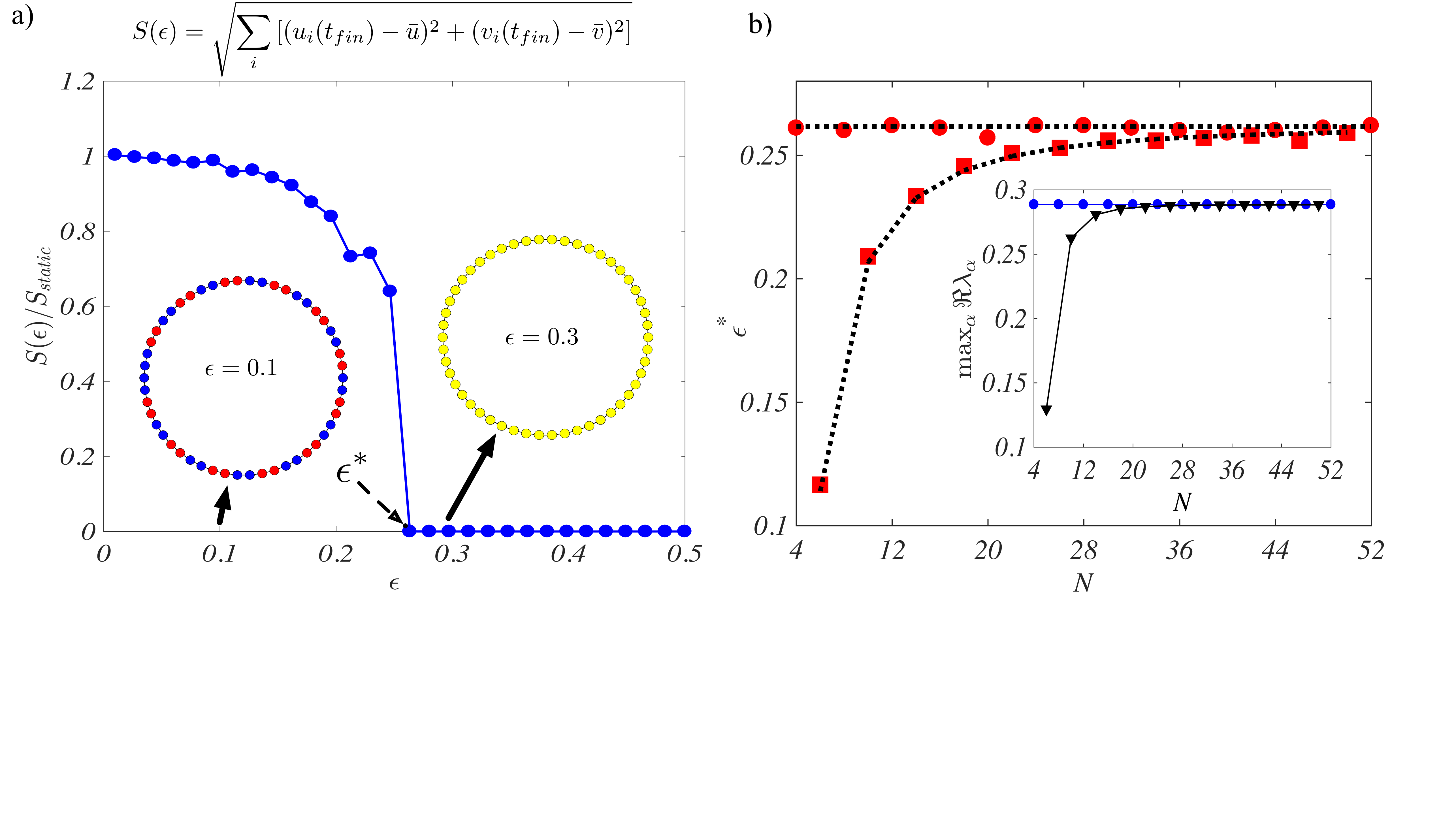}
\end{center}
\vspace{-2cm}\caption{The critical threshold $\epsilon^*$. Panel a): pattern amplitude, $S(\epsilon)$, as a function of $\epsilon$ normalised to the amplitude of the pattern for the averaged network 
$\langle \mathbf{A}\rangle=\gamma \mathbf{A}_1+(1-\gamma)\mathbf{A}_2$, for a $T$-periodic twin--network made of $N=50$ nodes. Insets: asymptotic patterns ($u$ variable) for $\epsilon=0.1<\epsilon^*$; blue (resp. red) nodes represent an high (resp. low) concentration of $u$, with respect to the homogeneous equilibrium. For $\epsilon=0.3>\epsilon^*$, the system converges toward the homogeneous equilibrium; nodes are plotted in yellow when the 
concentration of species $u$ is close to the homogeneous value. The employed reaction model is again the Brusselator and parameters are set as in Fig. \ref{fig:Fig1}.
Panel b): $\epsilon^*$ vs. $N$. The black dotted lines are drawn after equation~\eqref{eq:epsspect}, while the red symbols (circles for $N/2$ even and square for $N/2$ odd) are numerically computed. Inset: the maximum of the dispersion relation as a function of the network size (blue circles $N/2$ even and black down-triangles $N/2$ odd).}
\label{fig:Fig2}
\end{figure*}

We now introduce a periodically time varying network, specified by the following adjacency matrix $\mathbf{A}(t)$:  
\begin{equation}
\label{eq:At}
\mathbf{A}(t)=\begin{cases}
\mathbf{A}_1 &\text{if $\{t/T\}\in[0,\gamma)$}\\
\mathbf{A}_2 &\text{if $\{t/T\}\in[\gamma,1)$}\, ,
\end{cases}
\end{equation}
where $\{r\}$ denotes the fractional part of the real number $r$ and $\gamma\in(0,1)$ is the fraction of occurrence of the first network (and
thus $1-\gamma$ for the second one) during the period $T>0$. The Laplacian matrix $\mathbf{L}(t)$ can be computed accordingly. We get the averaged Laplacian matrix
%\begin{equation}
%\label{eq:aveLt}
$\langle \mathbf{L}\rangle = \gamma \mathbf{L}_1 +(1-\gamma)\mathbf{L}_2$
%\end{equation}
where $\mathbf{L}_i$ is the Laplacian matrix of the $i$--th network, $i=1,2$. As an example, assume  $\gamma=0.3$, $T=1$ and compute the dispersion relation associated
to the averaged network $\langle \mathbf{A}\rangle=\gamma \mathbf{A}_1 +(1-\gamma)\mathbf{A}_2$. 
From inspection of Fig.~\ref{fig:Fig1} panel c), it is immediately clear that $\Lambda^{\alpha}$, the
eigenvalues of the Laplacian $\langle \mathbf{L}\rangle$, fall in a region where $\Re\lambda_{\alpha}>0$. Hence, the Brussellator placed on the 
average network exhibits Turing patterns (see Fig.~\ref{fig:Fig1} panel d).
Based on the above,  we can find a positive $\epsilon^*$ such that for all $0<\epsilon<\epsilon^*$ the time dependent system~\eqref{eq:reac_diffacc} displays patterns. This is because 
this latter system is close enough to the averaged counterpart to be able to follow its orbits. On the other hand, no patterns can develop if $\epsilon$ is too large. These conclusions are clearly 
demonstrated in Fig.~\ref{fig:Fig2} panel a).

%\textbf{Determining the critical threshold $\epsilon^*$}
%\subsection{About the critical threshold $\epsilon^*$}
%\label{ssec:numstudepsstar}
We shall now complement the analysis by elaborating on a recipe to estimate the critical threshold $\epsilon^*$. Assume $\epsilon=1$ and linearise system (\ref{eq:reac_diffacc})
close to the stable homogeneous equilibrium $(u_i,v_i)=(\bar{u},\bar{v})$, for all $i=1,\dots,N$:
\begin{equation}
 \label{eq:mainlin}
\frac{d\delta X}{dt}=\mathbf{M}(t)\delta X\, ,
\end{equation}
where $\delta X=(u_1-\bar{u},\dots,u_N-\bar{u},v_1-\bar{v},\dots,v_N-\bar{v})^T$ and $\mathbf{M}(t)=\partial_X F(\bar{u},\bar{v})+\mathcal{L}(t)$, where $\partial_X F(\bar{u},\bar{v})$ is the Jacobian of the reaction part evaluated at the homogeneous equilibrium. $\mathbf{M}(t)$ is $T$-periodic and piecewise constant, more precisely $\mathbf{M}(t)=\partial_X F(\bar{u},\bar{v})+\mathcal{L}_1:=\mathbf{M}_1$ if $\{t/T\}\in[0,\gamma)$ and $\mathbf{M}(t)=\partial_X F(\bar{u},\bar{v})+\mathcal{L}_2:=\mathbf{M}_2$ if $\{t/T\}\in[\gamma,1)$.
We can thus solve exactly Eq.~\eqref{eq:mainlin} over one period
%\begin{equation}
 %\label{eq:mainlinsol1T}
${\delta X}(T)=e^{\mathbf{M}_2(1-\gamma)T}e^{\mathbf{M}_1\gamma T}\delta X(0)=:\mathbf{Q}\delta X(0)$
%\end{equation}
where the rightmost equality defines the {\em monodromy matrix}, $\mathbf{Q}$. Hence, for any $m\geq 1$,
%\begin{equation}
% \label{eq:mainlinsolmT}
${\delta X}(mT)=\mathbf{Q}^m\delta X(0)$ .
%\end{equation}
The stability of the system is therefore determined by the spectral radius of the monodromy matrix, $\rho(\mathbf{Q})$. If the spectrum is entirely contained in the unit disk,  $\delta X$ converges to zero 
and no patterns are allowed. Conversely, the applied perturbation can develop and eventually result in asymptotic stationary patterns. The same reasoning applies when $\epsilon$ is allowed to change freely. 
The monodromy matrix now reads $\mathbf{Q}_{\epsilon}=e^{\epsilon \mathbf{M}_2(1-\gamma)T}e^{\epsilon \mathbf{M}_1\gamma T}$ and one can conclude on the stability of the scrutinised system by computing its associated spectral radius $\rho(\mathbf{Q}_{\epsilon})$.  Assume, in line with the above, that $\rho(\mathbf{Q})=\rho(\mathbf{Q}_{\epsilon})\rvert_{\epsilon=1}<1$. Then, the critical threshold $\epsilon^*$ can be quantified as:
\begin{equation}
\label{eq:epsspect}
\epsilon^*=\min\{\epsilon>0: \forall s\geq \epsilon: \rho(\mathbf{Q}_{s})\leq 1\}\, .
\end{equation}

In Fig.~\ref{fig:Fig2}  (b) the predictions of the theory are challenged versus numerical simulations, performed for the twin network setting, as introduced above, for different values of the system size $N$. Excellent agreement is found which testifies on the correctness of the proposed interpretative scenario. Interestingly, $\epsilon^*$ returns a different profile in $N$, depending on whether $N/2$ is even or odd. The explanation resides in the corresponding magnitude of the dispersion relations, as depicted in the inset of  Fig.~\ref{fig:Fig2} (b). The instability is less pronounced when $N/2$ is odd, for small enough $N$. In this case, it is therefore necessary to push the modulation 
to the high frequencies even further to accompany the patterns crystallisation.  A closed approximate formula for $\epsilon^*$ can be also derived following an approach that is discussed in the SI. 

%This method returns:
%\begin{equation}
%\label{eq:epslow}
%\epsilon^*= \frac{1}{\Lambda_{12}^N\gamma(1-\gamma)T}\min \Big\{\frac{1}{D_u},\frac{1}{D_v}\Big\}\, ,
%\end{equation}
%where we set  $\Lambda_{12}^N=\max_{\alpha} \{\lvert \Lambda_{12}^{\alpha}\rvert\}$,  $\Lambda_{12}^{\alpha}$ being the eigenvalues of $\mathbf{L}_1-\mathbf{L}_2$. As shown in the SI, relation (\ref{eq:epslow}) 
%returns the correct order of magnitude for the critical threshold, but cannot account for the dependence in $N$, as outlined above. 

Summing up, patterns can be enforced in a reaction-diffusion system by switching periodically between distinct network configurations. Other examples aimed at illustrating our findings are discussed in the SI.  In the remaining part of this Letter we generalise the results along a few possible avenues of investigations, delegating technical details and representative applications to the annexed SI. 

%\textbf{Periodic, continuously time varying networks}
%\section{About $\epsilon^*$ for periodic, continuously time varying networks}
%\label{sec:anotherproof}

We can straightforwardly adapt the above reasoning to the case of a periodic, continuously time varying network and provide a reliable estimate for the critical threshold $\epsilon^*$.  In this  case $\mathbf{M}(t/\epsilon)=\partial_X F(\bar{u},\bar{v})+\mathcal{L}(t/\epsilon)$ is a matrix whose elements are continuous functions of time. By using the Floquet-Magnus expansion~\cite{Magnus1954,Casas2001,Blanes2009} one  can write 
%\begin{equation}
 %\label{eq:mainlinsol2acc}
${\delta X_{\epsilon}}(t)=e^{\mathbf{\Gamma}_{\epsilon}(t)}e^{t \mathbf{H}_{\epsilon}}\delta X_{\epsilon}(0)$,
%\end{equation}
where $\mathbf{\Gamma}_{\epsilon}(t)$ is $\epsilon T$-periodic. Writing
$\mathbf{H}_{\epsilon}=\sum_{n\geq 1}\mathbf{H}_{\epsilon, n}$, one can prove that $\mathbf{H}_{\epsilon,
  n}=\epsilon^{n-1}\mathbf{H}_{n}$. By evaluating the solution over one period $\epsilon T$ we obtain
\begin{equation}
 \label{eq:mainlinsol3acc}
{\delta X_{\epsilon}}(\epsilon T)=e^{\mathbf{\Gamma}_{\epsilon}(\epsilon T)}e^{\epsilon T \sum_{n\geq 1} \epsilon^{n-1}\mathbf{H}_{n}}\delta X_{\epsilon}(0)\, .
\end{equation}
Since $\mathbf{\Gamma}_{\epsilon}(\epsilon T)=0$, the stability of the homogeneous solution is determined by the eigenvalues of $\mathbf{H}_{\epsilon}$, the so-called characteristic exponents of the system. For sufficiently small $\epsilon$, one can write
$\mathbf{H}_{\epsilon}=\mathbf{H}_1+\mathcal{O}(\epsilon)$. The eigenvalues of $\mathbf{H}_{\epsilon}$ are hence close to those of $\mathbf{H}_1$. 
%In particular, the two sets of eigenvalues share the same sign of their respective real parts, since $\epsilon T$ is positive defined. 
Using the Magnus series one can determine 
\begin{equation}
 \label{eq:mainlinsol4acc}
\mathbf{H}_1=\frac{1}{T}\int_0^T \mathbf{M}(t)\, dt=\partial_X F(\bar{u},\bar{v}) +\langle \mathcal{L}\rangle \, ,
\end{equation}
which is nothing but the modified Jacobian matrix that applies to the averaged system. If $\epsilon$ is sufficiently small, the periodic, continuously time varying system behaves as the average one. Computing the spectrum of  (\ref{eq:mainlinsol4acc}) enables one to infer about the stability of the homogeneous solution. For a reaction-diffusion system hosted on network that is periodically modulated, with period $\epsilon T$ and $\epsilon<\epsilon^*$, patterns can arise, provided the average system can undergo a Turing like instability. Finally, we emphasise that the analysis readily extends to the case of non-periodically time varying networks. The only requirement is the existence of 
the generalised average $\langle\mathbf{L}\rangle=\lim_{T\rightarrow \infty}\frac{1}{T}\int_0^T \mathbf{L}(t) dt$. Relevant examples are addressed in the SI and include (i)  {\em quasi-periodic} time dependent networks and (ii) networks constructed using aperiodic infinite words from a binary alphabet, as the Sturmian words.

In conclusion, we have here shown that, by properly tuning the network topology over time, one can drive the emergence of self-organised patterns, reminiscent of the Turing instability. The reaction-diffusion system endowed with a time varying support behaves as its average analogue, provided the network dynamics is made sufficiently fast. Hence, patterns are obtained for a model attached to a network that changes over time, if they are predicted to occur on the associated average system. This is a novel route to pattern formation that we conjecture relevant for all those applications where different species interact via a  network that can adjust in time, as follows either an endogenous or exogenous drive.

\textit{Acknowledgments}
The work of T.C. presents research results of the Belgian Network DYSCO (Dynamical Systems, Control, and Optimization), funded by the Interuniversity Attraction Poles Programme, initiated by the Belgian State, Science Policy Office. The scientific responsibility rests with its author(s). D.F.  acknowledges financial support from H2020-MSCA-ITN-2015 project COSMOS 642563.

\appendix
\section{SUPPLEMENTARY INFORMATIONS}
\section{Proof of the main result}
\label{sec:proof}

The aim of this section is to provide a detailed derivation of our main result that we will cast in the form of a theorem. 
Let us first recall the definition of the model and then proceed to prove the theorem.

We assume, for {the} sake of simplicity, two different species that live  and interact on a periodically time varying network. The number of nodes $N$ is kept fixed, while pairwise weighted links adjust in time. The network structure is embedded in a periodic time varying weighted adjacency matrix $A_{ij}(t)$, that results in time varying link strength, $s_i(t)=\sum_j A_{ij}(t)$. Further, we introduce the time varying Laplacian matrix defined $L_{ij}(t)=A_{ij}(t)-s_i(t)\delta_{ij}$. Observe that for all $t$ one has $\sum_j L_{ij}(t)=0$, namely $\mathbf{L}(t)$ is a row stochastic matrix.

Species diffuse across the network and interact on the same node as dictated by specific non-linear reaction terms. The model (equations (1) in the main body of the paper) reads:
\begin{eqnarray}
\dot{u}_i&=& f(u_i,v_i) + D_u\sum_{j=1}^{N}{L}_{ij}(t/\epsilon) u_j \nonumber\\
\dot{v}_i&=& g(u_i,v_i) + D_v\sum_{j=1}^{N}L_{ij}(t/\epsilon) v_j
\label{eq:reac_diff}
\end{eqnarray}
where $f$ and $g$ are non-linear fucntions and $D_u$ and $D_v$ denote the the diffusion coefficients associated, respectively, to species $u$ and $v$. The parameter $\epsilon$ sets the rate of the network evolution. For a subsequent use let us rewrite the above system in a more compact form by defining the $2N$--dimensional vector $\vec{x}=(u_1,\dots,u_N,v_1,\dots,v_N)$:
\begin{equation}
\dot{\vec{x}}(t)= F(\vec{x}) +\mathcal{L}(t/\epsilon) \vec{x}\, ,
\label{eq:reac_diffacc3}
\end{equation}
where $F(\vec{x})=(f(u_1,v_1),\dots,f(u_N,v_N),g(u_1,v_1),\dots,g(u_N,v_N))^T$ and 
\begin{equation}
\mathcal{L}(t) \vec{x}:=\left(
\begin{smallmatrix}
 D_u \mathbf{L}(t) & 0 \\ 0 &  D_v \mathbf{L}(t) 
 \end{smallmatrix}
\right)\vec{x}\, .
\label{eq:DLx}
\end{equation}

To study the emergence of patterns for model ~\eqref{eq:reac_diff}, as follows a symmetry breaking instability of the Turing type, one cannot invoke the standard machineries, as eigenvectors and eigenvalues depend on time.
To compute the dispersion relation following the canonical approach additional assumptions are to be enforced, as e.g. discussed in ~\cite{Boccaletti2006}. To avoid this, we proceed with an alternative route which involves 
dealing with the so called {\it theorem of averaging} hereafter recalled for consistency (the interested reader can consult e.g.~\cite{verhulst} for further details on this topic):

\begin{theorem}[Averaging in the periodic case]
\label{thm:avper}
Let us consider the system
\begin{equation}
 \label{eq:thesystem}
\dot{x}=\epsilon f(t,x)\, ,\quad x(0)=x_0\in D\subset\mathbb{R}^n\, ,\end{equation}
where $f$ and $\partial_x f$ are defined, continuous and bounded in $[0,\infty)\times \mathbb{R}^n$ and assume $f(t,\cdot)$ to be $T$-periodic.

Let $\langle f\rangle (y)$ be the time average of $f(t,y)$, that is
\begin{equation}
 \label{eq:tavgen}
 \langle f\rangle (y)=\frac{1}{T}\int_0^T f(t,y)\, dt\, ,
\end{equation}
and let $y(t)$ be the solution of 
\begin{equation*}
\dot{y}=\epsilon \langle f\rangle (y)\, ,\quad y(0)=x_0\in D\subset\mathbb{R}^n\, .
\end{equation*}

Then $x(t)-y(t)=\mathcal{O}(\epsilon)$ for $t=\mathcal{O}(1/\epsilon)$.
\end{theorem}

Based on the theorem above, and as discussed in the main body of the paper, we can relate the behaviour of system~\eqref{eq:reac_diff}  to the behaviour of its {\em averaged} homologue, defined by replacing $\mathbf{L}(t)$ with the time averaged operator $\langle\mathbf{L}\rangle$ defined as:

\begin{equation}
 \label{eq:ave1}
\langle\mathbf{L}\rangle=\frac{1}{T}\int_0^T \mathbf{L}(t)\, dt\, ,
\end{equation}
where $T>0$ is the period. As a side remark, we anticipate that our result holds true also if the time evolution of the network is not periodic, provided a generalised average exists, namely:
\begin{equation*}
\langle\mathbf{L}\rangle=\lim_{T\rightarrow \infty}\frac{1}{T}\int_0^T \mathbf{L}(t)\, dt\, .
\end{equation*}

More precisely we can state and proof the following theorem
\begin{theorem}
\label{thm:mainthm}
 Let us define the averaged system~\footnote{$F(\vec{x})$ being  time independent, the averaging affects only on the Laplacian matrix.}:
 %\begin{eqnarray}
%\dot{u}_i(t)&=& f(u_i,v_i) + D_u\sum_{j=1}^{N}\langle L_{ij}\rangle u_j \nonumber\\
%\dot{v}_i(t)&=& g(u_i,v_i) + D_v\sum_{j=1}^{N}\langle L_{ij}\rangle v_j
%\label{eq:reac_diffave}
%\end{eqnarray}
%or in compact form
\begin{equation}
\dot{\vec{x}}(t)= F(\vec{x}) +\langle\mathcal{L}\rangle \vec{x}\, ,\quad \vec{x}(0)=\vec{x}_0\in \mathbb{R}^{2N},
\label{eq:reac_diffacc3b}
\end{equation}
where $\vec{x}=(u_1,\dots,u_N,v_1,\dots,v_N)$, $\langle\mathcal{L}\rangle=\left(
\begin{smallmatrix}
 D_u \langle\mathbf{L}\rangle & 0 \\ 0 &  D_v \langle\mathbf{L}\rangle
 \end{smallmatrix}
\right)$ and $\langle\mathbf{L}\rangle$ is defined using Eq.~\eqref{eq:ave1}. Assume moreover that the above system exhibits standard Turing patterns, namely that a stable homogeneous equilibrium $\bar{x}=(\bar{u},\dots,\bar{u},\bar{v},\dots,\bar{v})$ of system (\ref{eq:reac_diffacc3b}) can turn unstable, for an appropriate choice of the parameters involved, upon application of a non homogeneous perturbation. 

Then there exists $\epsilon^*>0$ such that the {\em fast varying } version of Eq.~\eqref{eq:reac_diffacc3}
%Eq.~\eqref{eq:reac_diff}, namely
%\begin{eqnarray}
%\dot{u}_i(t)&=& f(u_i,v_i) + D_u\sum_{j=1}^{N}{L}_{ij}(t/\epsilon) u_j(t) \nonumber\\
%\dot{v}_i(t)&=& g(u_i,v_i) + D_v\sum_{j=1}^{N}L_{ij}(t/\epsilon) v_j(t)
%\label{eq:reac_diffacc}
%\end{eqnarray}
%or in compact form
\begin{equation}
\dot{\vec{x}}(t)= F(\vec{x}) +\mathcal{L}(t/\epsilon) \vec{x}\, ,
\label{eq:reac_diffacc3acc}
\end{equation}
displays patterns of the Turing type for all $0<\epsilon<\epsilon^*$.
\end{theorem}

\textbf{Proof}
Let us first introduce the rescaled time $\tau=t/\epsilon$ and the adapted variables $\vec{w}(\tau)=\vec{x}(t)\rvert_{t=\epsilon \tau}$, to rewrite Eq.~\eqref{eq:reac_diffacc3acc} as (with the notation $^{\prime}=d/d\tau$):
\begin{equation}
{\vec{w}}^\prime(\tau)= \epsilon\left[F(\vec{w}) +\mathcal{L}(\tau) \vec{w}\right]\, .
\label{eq:reac_diffacc3accw}
\end{equation}

We can then apply theorem~\ref{thm:avper} and obtain
\begin{equation*}
\vec{w}(\tau)-\vec{y}(\tau)=\mathcal{O}(\epsilon)\, ,
\end{equation*}
for all $0<\epsilon <\epsilon^*$, where $\vec{y}(\tau)$ is the solution of the averaged system
\begin{equation}
\label{eq:reac_diffaccy}
\vec{y}\,^{\prime}(\tau)=\epsilon\left[F(\vec{y}) +\langle\mathcal{L}\rangle \vec{y}\right]\, ,\quad \vec{y}(0)=\vec{w}(0)\in \mathbb{R}^{2N}\, .
\end{equation}

Back to the original variables, $\vec{x}$ and time $t$,  the last statement can be rephrased as follows. Let $\vec{x}$ be the solution of Eq.~\eqref{eq:reac_diffacc3acc} and $\vec{z}(t)=\vec{y}(t/\epsilon)$ the solution of Eq.~\eqref{eq:reac_diffacc3b} with the same initial conditions, then 
 \begin{equation*}
\vec{x}(t)-\vec{z}(t)=\mathcal{O}(\epsilon)\text{ for all $0<\epsilon<\epsilon^*$ and $t=\mathcal{O}(1)$.}
\end{equation*}
Namely the two above orbits stay close over a macroscopic time window. Hence, if a symmetry breaking instability occurs for system 
~\eqref{eq:reac_diffacc3b}, paving the way to the subsequent pattern derive, the same holds 
for the original system~\eqref{eq:reac_diffacc3acc}.

\section{On the estimate of the critical threshold}

The aim of this section is to provide a detailed derivation of the formulae employed in the main text to estimate $\epsilon^*$. To this end
we consider again the reference system written in a compact vectorial form: 
\begin{equation}
\label{eq:system_original_vectorform}
\dot{\vec{x}}(t)= F(\vec{x}) +\mathcal{L}(t/\epsilon) \vec{x}\, ,
\end{equation}
where $\vec{x}=(u_1,\dots,u_N,v_1,\dots,v_N)^T$, $F(\vec{x})=(f(u_1,v_1),\dots,f(u_N,v_N),g(u_1,v_1),\dots,g(u_N,v_N))^T$ and $\mathcal{L}(t)=\left(
\begin{smallmatrix}
 D_u \mathbf{L}(t) & 0 \\ 0 &  D_v \mathbf{L}(t) 
 \end{smallmatrix}
\right)$.

Assume there exists $\bar{u},\bar{v}$ such that $\bar{x}=(\bar{u},\dots,\bar{u},\bar{v},\dots,\bar{v})^{T}$ is an homogeneous equilibrium of~\eqref{eq:system_original_vectorform}, and write $\delta \vec{x}=\vec{x}-\bar{x}$ to denote a tiny perturbation. Linearising system (\ref{eq:system_original_vectorform}) close to the homogeneous solution yields:
\begin{equation}
\dot{\delta \vec{x}}=\mathbf{M}_{\epsilon}(t)\delta \vec{x}\qquad\mbox{with}\qquad \mathbf{M}_{\epsilon}(t)=\partial_{x}F(\bar{x})+\mathcal{L}(t/{\epsilon})\, ,
\label{eq:system_accelerated_vectorform-linearized}
\end{equation}
where $\partial_{x}F(\bar{x})$ is the Jacobian matrix of $F$ evaluated at $\bar{x}$.

In the rest of the section we shall assume that no Turing instability is allowed for $\epsilon=1$. In other words, the imposed perturbation fades away when 
$\epsilon$ is set equal to $1$, and system (\ref{eq:system_original_vectorform}) converges to its  homogeneous equilibrium. At variance, 
the time averaged system, which corresponds to
the limiting case $\epsilon\approx0$, can exhibit patterns of the Turing class. In the following, we shall write for short $\mathbf{M}_{\epsilon=1}=\mathbf{M}$. Our goal is to compute, in this setting, the critical threshold for the control parameter 
$\epsilon$ below which the fast varying version of the inspected system can also display Turing like patterns.

\subsection{Temporal network of contact sequences
\label{sub:Switching-networks}}

Let us consider a finite collection of networks indexed by $k\in\mathcal{K}$ and label $\left\{ \mathbf{L}^{[k]}:k\in\mathcal{K}\right\} $ their associated Laplacian matrices. For fixed reaction terms, we then define, for every $k\in\mathcal{K}$, the matrix $\mathbf{M}^{[k]}=\partial_{x}F(\bar{x})+\mathcal{L}^{[k]}$, where  $\mathcal{L}^{[k]}=\left(
\begin{smallmatrix}
 D_u \mathbf{L}^{[k]} & 0 \\ 0 &  D_v \mathbf{L}^{[k]} 
 \end{smallmatrix}
\right)$. We further assume that for every linearised system involving a given $\mathbf{M}^{[k]}$, i.e. when the index $k$ is kept fixed,  the zero solution is stable with respect to injection of a small non homogeneous perturbation $\delta \vec{x}$.

Let us now introduce a switching signal, $\sigma(t)$, that is, a piecewise constant function $\sigma:[0,\infty)\rightarrow\mathcal{K}$. This provides a direct map between time and indexes, thus enabling us to associate to each time its corresponding network
\begin{equation}
\mathcal{L}(t)=\mathcal{L}^{[\sigma(t)]}\, . 
\end{equation}
Working along these lines, we generalise the construction of twin networks as discussed in the main body of the paper. For the twin networks, in fact,  $\mathcal{K}=\{1,2\}$ and $\sigma(t)=1$ if $\mathrm{mod}(t,T)/T\in[0,\gamma)$ and $\sigma(t)=2$ otherwise. We are thus interested in the stability of the $1$-parameter family of linearised systems
\begin{equation}
\dot{\delta \vec{x}}=\mathbf{M}_\epsilon^{[\sigma(t)]}\delta \vec{x}\qquad\mbox{with}\qquad \mathbf{M}_\epsilon^{[\sigma(t)]}=\partial_{x}F(\bar{x})+\mathcal{L}^{[\sigma(t/\epsilon)]}\,.
\label{eq:system_accelerated_vectorform-linearizedk}
\end{equation}
If the interval between any two consecutive discontinuities of $\sigma(t)$ is (on average)
above a given threshold $\tau_{D}$, called the dwell time of the system, then the zero solution of Eq.~\eqref{eq:system_accelerated_vectorform-linearizedk}
is stable (see for instance \cite{Morse_1996_dwell_time}). 

Let $0=t_{0}<t_{1}<t_{2}<\dots$ be the times of discontinuities of $\sigma(t)$, and define $\tau_{k}$ by $t_{k}+\tau_{k+1}=t_{k+1}$ for $k\in\mathbb{N}$. The switching signal $\sigma(t)$ is constant over every interval $[t_{k},t_{k+1})$ and therefore the solution of Eq.~\eqref{eq:system_accelerated_vectorform-linearizedk} can be explicitly computed as 
\begin{equation}
\delta \vec{x}(\epsilon t_{k+1})=\exp(\epsilon\tau_{k+1}\mathbf{M}_\epsilon^{[\sigma(\epsilon t_{k})]})\delta \vec{x}(\epsilon t_{k}),\quad k\in\mathbb{N}\, .
\label{eq:system_original_linearized2}
\end{equation}

We then get a discrete time system that is stable under injection of a small non homogeneous perturbation, if the joint spectral radius satisfies:
\begin{equation}
\rho\left\{ \exp(\epsilon \tau_k \mathbf{M}_{\epsilon}^{[\sigma(\epsilon t_{k})]}),\ k\in\mathbb{N})\right\} <1\, .
\end{equation}
Observe that even for a set composed by only two matrices, the joint spectral radius is cumberosme to compute \cite{Vankeerberghen2014_JSR}. To allow a more straightforward computation of the critical threshold $\epsilon^*$, we proceed by making the simplifying assumption that the  switching signal $\sigma(t)$ has a finite number of discontinuities $(n)$ and it is $T-$periodic, with $T=\sum_{i=1}^{n}\tau_{i}$. 

As a result, we have that the solution of \eqref{eq:system_original_linearized2} at every integer multiple of periods  $m \epsilon T$ is given by 
\begin{equation}
\delta \vec{x}(m \epsilon T)=\left(\mathbf{Q}_{\epsilon}(\epsilon T)\right)^{m}\delta \vec{x}(0)\, ,
\end{equation}
where the time-evolution operator over one period, 
\begin{equation*}
\mathbf{Q}_{\epsilon}(\epsilon T) =\prod_{k=1}^{n}\exp(\epsilon\tau_{k}\mathbf{M}_{\epsilon}^{[\sigma(\epsilon t_{k-1})]})=\prod_{k=1}^{n}\exp(\epsilon\tau_{k}\mathbf{M}^{[\sigma(t_{k-1})]})\, ,
\end{equation*}
where we have used that 
%$M(t)=M^{\left[\sigma(t)\right]}$ and $M_{\epsilon}(t)=M(t/\epsilon)$.
$\mathbf{M}_{\epsilon}^{\left[\sigma(t)\right]}=\mathbf{M}^{\left[\sigma(t/\epsilon)\right]}$ and thus $\mathbf{M}_{\epsilon}^{\left[\sigma(\epsilon t)\right]}=\mathbf{M}^{\left[\sigma(t)\right]}$ (see Eq.~\eqref{eq:system_accelerated_vectorform-linearizedk}).

The critical value of the acceleration parameter $\epsilon$ is thus given by 
\begin{equation}
\epsilon_*=\min\left\{ \epsilon>0\ :\ \forall s\geq \epsilon,\rho(\mathbf{Q}_{s}(s T))\leq1\right\} \, ,\label{eq:critical_epsilon_piecewise_constant}
\end{equation}
where $\rho$ denotes the spectral radius.

\subsection{Continuously varying networks\label{sub:continuously-varying-networks}}

We now consider the more general case of a continuously time varying network,
with Laplacian matrix $\mathbf{L}(t)$ whose entries are continuous integrable functions
of time. The solution of the linearised system with $\epsilon=1$, 
\begin{equation}
\dot{\delta \vec{x}}=\mathbf{M}(t)\delta \vec{x}\qquad\mbox{with}\qquad \mathbf{M}(t)=\partial_{x}F(\bar{x})+\mathcal{L}(t)\, ,\label{eq:system_original_linearized-1}
\end{equation}
can be written as
\begin{equation}
\delta \vec{x}(t)=\exp\left(\int_{0}^{t}\mathbf{M}(t)\, dt\right)\delta \vec{x}(0)\, ,
\label{eq:solution-commuting-matrices}
\end{equation}
if and only if the matrices $\mathbf{M}(t_{1})$ and $\mathbf{M}(t_{2})$ commute for any pair $t_{1},t_{2}$. Let us observe that, as $\mathbf{M}$ contains both a term related to the reactions and another related to the diffusion, the above assumption may  not be satisfied, even if the Laplacian matrices associated to the networks do commute. The commutation between the Laplacian matrix and the reaction part is also needed. However one can always write: 
\begin{equation}
\delta \vec{x}(t)=\exp\left(\mathbf{\Omega}(t)\right)\delta \vec{x}(0)\, ,
\label{eq:delta-X-is-exp-Omega}
\end{equation}
where $\mathbf{\Omega}(t)=\sum_{n=1}^{\infty}\mathbf{\Omega}_{n}(t)$, is known as the Magnus expansion
\cite{Magnus_1954,Blanes2009_expansion-and-some-applications}.
From 
\begin{equation}
\frac{\mathrm{d}\mathbf{\Omega}(t)}{\mathrm{d}t}=\sum_{n=0}^{\infty}\frac{B_{n}}{n!}\mathrm{ad}_{\Omega}^{n}\mathbf{A}\, ,
\end{equation}
where the $B_{n}$'s are the first Bernoulli numbers ($B_{1}=-1/2$), $\mathrm{ad}_{X}\mathbf{A}=[\mathbf{X},\mathbf{A}]=\mathbf{X}\mathbf{A}-\mathbf{A}\mathbf{X}$ denote the matrix commutator and  $\mathrm{ad}_{X}^{0}\mathbf{A}=\mathbf{A}$, $\mathrm{ad}_{X}^{k}\mathbf{A}=[\mathbf{X},\mathrm{ad}_{X}^{k-1}\mathbf{A}]$, one can iteratively compute all the terms of the series. For example, 
\begin{align}
\mathbf{\Omega}_{1}(t) & =\int_{0}^{t}\mathbf{M}(t_{1})\mathrm{d}t_{1}\\
\mathbf{\Omega}_{2}(t) & =\frac{1}{2}\int_{0}^{t}\int_{0}^{t_{1}}\left[\mathbf{M}(t_{1}),\mathbf{M}(t_{2})\right]\mathrm{d}t_{2}\mathrm{d}t_{1}\\
\mathbf{\Omega}_{3}(t) & =\frac{1}{6}\int_{0}^{t}\int_{0}^{t_{1}}\int_{0}^{t_{2}}\left(\left[\mathbf{M}(t_{1}),\left[\mathbf{M}(t_{2}),\mathbf{M}(t_{3})\right]\right]+\left[\mathbf{M}(t_{3}),\left[\mathbf{M}(t_{2}),\mathbf{M}(t_{1})\right]\right]\right)\mathrm{d}t_{3}\mathrm{d}t_{2}\mathrm{d}t_{1}.
\end{align}

One can also prove~\cite{Moan_2008-convergence-magnus-series} the convergence of the above series provided $\int_{0}^{t}\left\Vert \mathbf{M}(\tau)\right\Vert _{2}\mathrm{d}\tau<\pi$.

As previously done when dealing with the switching network case, we assume for simplicity that the network evolution is periodic. Hence, the entries of $\mathbf{L}(t)$ are $T-$periodic continuous functions. The Floquet theorem ensures thus that the solution of \eqref{eq:system_original_linearized-1} can be written as 
\begin{equation}
\delta \vec{x}(t)=\mathbf{P}(t)\exp\left(t\mathbf{F}\right)\delta \vec{x}(0)\, ,
\end{equation}
where $\mathbf{P}$ is $T$-periodic and bounded, and $\mathbf{F}$ does not depend on time. The asymptotic stability of the null solution is determined by the eigenvalues of $\mathbf{F}$, known as the characteristic exponents of the system. The Floquet-Magnus expansion~\cite{Casas2001_floquet-exponential-perturbative} allows one to write 
\begin{equation}
\delta \vec{x}(t)=\exp\left(\mathbf{\Lambda}(t)\right)\exp\left(t\mathbf{F}\right)\delta \vec{x}(0)\, ,
\end{equation}
where the involved matrices are obtained as the sums of series~\footnote{The convergence for the series for $F$ follows from the above quoted condition $\int_{0}^{T}\left\Vert \mathbf{M}(\tau)\right\Vert _{2}\,d\tau<\pi$,
whereas for the series related to $\mathbf{\Lambda}(t)$ we have the more restrictive
sufficient condition $\int_{0}^{T}\left\Vert \mathbf{M}(\tau)\right\Vert _{2}\,d\tau<0.20925$.}
\begin{equation}
\mathbf{\Lambda}(t)=\sum_{n=1}^{\infty}\mathbf{\Lambda}_{n}(t),\quad \mathbf{F}=\sum_{n=1}^{\infty}\mathbf{F}_{n}\, .
\end{equation}
The matrix $\mathbf{\Lambda}(t)$ is $T$-periodic, with $\mathbf{\Lambda}(0)=\mathcal{O}$. Since $\delta \vec{x}(T)=\exp[\mathbf{\Omega}(T)]\delta \vec{x}(0),$ we have the identity:
\begin{equation}
\mathbf{F}_{n}=\mathbf{\Omega}_{n}(T)/T,\quad n\geq1\, .
\end{equation}

Let us now turn to considering the system \eqref{eq:system_accelerated_vectorform-linearized} in the case $\epsilon<1$.
We write $\mathbf{\Omega}_{\epsilon}(t)$ for the Magnus expansion corresponding
to $\mathbf{M}_{\epsilon}(t)=\mathbf{M}(t/\epsilon)$, and apply the same notation
to $\mathbf{\Lambda}(t)$ and $\mathbf{F}$. This leads to 
\begin{equation}
\delta \vec{x}(t)=\exp\left(\sum_{n=1}^{\infty}\mathbf{\Lambda}_{\epsilon,n}(t)\right)\exp\left(t\sum_{n=1}^{\infty}\mathbf{F}_{\epsilon,n}\right)\delta \vec{x}(0)\, ,
\end{equation}
where $\mathbf{\Lambda}_{\epsilon}(t)$ is $\epsilon T$-periodic, and
\begin{equation}
\mathbf{F}_{\epsilon,n}=\mathbf{\Omega}_{\epsilon,n}(\epsilon T)/(\epsilon T),\quad n\geq1.
\end{equation}
The stability is determined by the eigenvalues of the characteristic exponent
$\epsilon T\mathbf{F}_{\epsilon}$, or equivalently, by looking at the
sign of the real part of the eigenvalues of $\mathbf{\Omega}_{\epsilon}(\epsilon T)/(\epsilon T)$.

Let us observe that for every $n\geq1$, we have $\mathbf{\Omega}_{\epsilon,n}(\epsilon t)=\epsilon^{n}\mathbf{\Omega}_{n}(t)$. In conclusion we obtain
\begin{equation}
\mathbf{\Omega}_{\epsilon}(\epsilon T)=\sum_{n=1}^{\infty}\epsilon^{n}\mathbf{\Omega}_n(T)\, ,
\label{eq:Omega-epsilon-T}
\end{equation}
and, as a consequence, the stability of the zero solution of the fast-varying system ($\epsilon<1$) is determined by the sign of the real part of the eigenvalues
of $\sum_{n=1}^{\infty}\epsilon^{n-1}\mathbf{\Omega}(T)/T$. 

As remarked when discussing the implication of the averaging theorem, in the limiting case $\epsilon\rightarrow0$, the averaged network can be used to determine the stability of the system. Indeed,
if $\epsilon>0$ is small enough, the eigenvalues of $\sum_{n=1}^{\infty}\epsilon^{n-1}\mathbf{\Omega}_{n}(T)/T=\mathbf{\Omega}_{1}/T+o(\epsilon)$
lie on the same portion of the imaginary axis as the eigenvalues of 
\begin{equation}
\frac{\mathbf{\Omega}_{1}}{T}=\frac{1}{T}\int_{0}^{T}\mathbf{M}(t)\, dt=\partial_{x}{F}(\bar{x})+\langle \mathcal{L}(t)\rangle .
\end{equation}

The critical value of $\epsilon$ can be computed by imposing 
\begin{equation}
\epsilon^{*}=\min\left\{ \epsilon>0\ :\ \gamma>\epsilon\implies\max_{\lambda\in\sigma_{\gamma}}\Re\lambda\leq0\right\} \label{eq:critical-epsilon-regular-magnus}
\end{equation}
where $\sigma_{\gamma}$ is the spectrum of \eqref{eq:Omega-epsilon-T}
for $\epsilon=\gamma$. 

The computation of the first terms of the Magnus expansion, truncated as desired so as
to reach the necessary precision on $\epsilon^{*}$, can be carried out by
using the recursive formulas first obtained in \cite{Klarsfled_1989_recursive_terms}:
\begin{equation}
\mathbf{\Omega}_{n}=\sum_{j=0}^{n-1}\frac{B_{j}}{j!}\int_{0}^{t}\mathbf{S}_{n}^{(j)}(t_{1})\mathrm{d}t_{1},\quad n\geq1\label{eq:Omega_n_recursive}
\end{equation}
where the $\mathbf{S}_{n}^{(j)}$ follow from the recursion
\begin{alignat}{1}
\mathbf{S}_{n}^{(j)} & =\sum_{m=1}^{n-j}\left[\mathbf{\Omega}_{m},\mathbf{S}_{n-m}^{(j-1)}\right],\quad1\leq j\leq n-1\\
\mathbf{S}_{1}^{(0)} & =\mathbf{M}(t)\\
\mathbf{S}_{n}^{(0)} & =0,\quad n>1.\label{eq:Sn0}
\end{alignat}
Working out these formulas explicitly, 
\begin{equation}
\mathbf{\Omega}_{n}(t)=\sum_{j=1}^{n-1}\frac{B_{j}}{j!}\sum_{\stackrel{k_{1}+\ldots+k_{j}=n-1}{k_{1}\geq0,\ldots,k_{j}\geq0}}\int_{0}^{t}\mathrm{ad}_{\Omega_{k_{1}}(\tau)}\mathrm{ad}_{\Omega_{k_{2}}(\tau)}\ldots\mathrm{ad}_{\Omega_{k_{j}}(\tau)}\mathbf{M}(\tau)\mathrm{d}\tau,\quad n\geq2.\label{eq:Omega_n_worked-out}
\end{equation}
The above formulas are straightforward to implement, but prove numerically heavy to calculate
since they involve multiple integrals of nested commutators. To overcome this limitation one can use the {\em Time-stepping} method which consists in performing a partition of the interval $[0,T]$
in $m$ subintervals of length $h$ and then compute the Taylor series for $\mathbf{M}(t)$ using the midpoint $t^{[k]}=(k-1)h+\frac{h}{2}=(k-\frac{1}{2})h$:
\begin{equation}
\mathbf{M}(t)=\sum_{j=0}^{\infty}\mathbf{m}_{j}^{\left[k\right]}(t-t^{\left[k\right]})^{j},\ t_{k-1}\leq t<t_{k},\quad\mbox{with}\quad \mathbf{m}_{j}^{\left[k\right]}=\frac{1}{j!}\left.\frac{\mathrm{d}^{j}\mathbf{M}(t)}{\mathrm{d}t^{j}}\right|_{t=t^{\left[k\right]}}.
\end{equation}
Inserting this relation into the recursive formulas given above yields~\cite{Blanes2000_high-order},
up to order $5$:
\begin{align}
\mathbf{\Omega}_{1}^{\left[k\right]} & =h\mathbf{m}_{0}^{\left[k\right]}+h^{3}\frac{1}{12}\mathbf{m}_{2}^{\left[k\right]}+h^{5}\frac{1}{80}\mathbf{m}_{4}^{\left[k\right]}+o(h^{7})\\
\mathbf{\Omega}_{2}^{\left[k\right]} & =h^{3}\frac{-1}{12}\left[\mathbf{m}_{0}^{\left[k\right]},\mathbf{m}_{1}^{\left[k\right]}\right]+h^{5}\left(\frac{-1}{80}\left[\mathbf{m}_{0}^{\left[k\right]},\mathbf{m}_{3}^{[k]}\right]+\frac{1}{240}\left[\mathbf{m}_{1}^{\left[k\right]},\mathbf{m}_{2}^{\left[k\right]}\right]\right)+o(h^{7})\\
\mathbf{\Omega}_{3}^{\left[k\right]} & =h^{5}\left(\frac{1}{360}\left[\mathbf{m}_{0}^{\left[k\right]},\mathbf{m}_{0}^{\left[k\right]},\mathbf{m}_{2}^{\left[k\right]}\right]-\frac{1}{240}\left[\mathbf{m}_{1}^{\left[k\right]},\mathbf{m}_{0}^{\left[k\right]},\mathbf{m}_{1}^{\left[k\right]}\right]\right)+o(h^{7})\\
\mathbf{\Omega}_{4}^{\left[k\right]} & =h^{5}\frac{1}{720}\left[\mathbf{m}_{0}^{\left[k\right]},\mathbf{m}_{0}^{\left[k\right]},\mathbf{m}_{0}^{\left[k\right]},\mathbf{m}_{1}^{\left[k\right]}\right]+o(h^{7}).
\end{align}

Here, we have used the simplified notation $[x_{1},x_{2},\ldots,x_{j}]=[x_{1},[x_{2},[\ldots,[x_{j-1},x_{j}]\ldots]]]$.
Over each time interval $[t_{k-1},t_{k})$, we get as a viable approximation
to the time evolution operator from $\delta \vec{x}(t_{k-1})$ to $\delta \vec{x}(t_{k})$, relative to the reference case $\epsilon=1$ : 
\begin{equation}
\exp\left(\mathbf{\Omega}(t_{k},t_{k-1})\right)=\exp\left(\sum_{n=1}^{4}\mathbf{\Omega}_{n}^{\left[k\right]}+o(h^{7})\right)
\end{equation}
After appropriate truncation of the series, we have
\begin{equation}
\delta \vec{x}(T)=\prod_{k=1}^{m}\exp\left(\sum_{n\geq1}\mathbf{\Omega}_{n}^{\left[k\right]}\right)\delta \vec{x}(0)\, .
\end{equation}
Note that here no integration of $\mathbf{M}(t)$ is required, so the computation is much faster to handle numerically. The analysis can be readily extended to the setting $\epsilon<1$. If one 
considers again $m$ subintervals~\footnote{Let us note that if $m$ is large enough, the convergence of the Magnus series on each subinterval is guaranteed.} in $[0,\epsilon T]$, we then have
\begin{equation}
\delta \vec{x}(\epsilon T) =\prod_{k=1}^{m}\exp\left(\sum_{n\geq1}\mathbf{\Omega}_{\epsilon,n}^{\left[k\right]}(\epsilon t_{k},\epsilon t_{k-1})\right)\delta \vec{x}(0)=\prod_{k=1}^{m}\exp\left(\sum_{n\geq1}\epsilon^{n}\mathbf{\Omega}_{n}^{\left[k\right]}(t_{k},t_{k-1})\right)\delta \vec{x}(0)\,.\label{eq:time-evolution-operator-accelerated-time-stepping}
\end{equation}
As before, we only need to compute the terms of the Magnus series
for the original system, that we then multiply by the appropriate
integer power of $\epsilon$. In doing so, we can obtain an estimate for the critical value $\epsilon^*$ by evaluating the spectral radius of the matrix that appears in the right hand side of the latter equation.

\subsection{A closed-form approximate expression for $\epsilon^*$}
\label{sec:epsilonstar}

The aim of this section is to present an alternative method, directly inspired by the proof of the averaging theorem, to compute the critical threshold $\epsilon^*$. This procedure avoids dealing with the computation of the 
monodromy matrix and returns a closed formula for $\epsilon^*$, which proves adequate versus numerical simulations.

Digging into the proof of the {averaging} theorem one can appreciate that it relies on an {\em invertible change of coordinates} matching the two systems (Eq.~\eqref{eq:reac_diffacc3accw} and Eq.~\eqref{eq:reac_diffaccy}). Our idea is hence to quantify $\epsilon^*$ by determining the range of $\epsilon$ for which the sought invertibility can be achieved. The required change of variables is given by
\begin{equation}
  \label{eq:changevar}
  \vec{w}=\vec{y}+\epsilon \vec{u}(\tau,\vec{y})\, ,
\end{equation}
where
$\vec{u}$ is explicitely given by
\begin{equation}
\label{eq:uy}
  \vec{u}(s,\vec{y})=\int_0^\tau\left[\mathcal{L}(r)\vec{y}-\langle\mathcal{L}\rangle\vec{y}\right]\,
  dr\, ,
\end{equation}
where the terms involving the reaction part, $F$ is Eq.~\eqref{eq:system_original_vectorform}, cancel out because they do not depend explicitly on time.

To be able to invert relation~\eqref{eq:changevar} one has
to require that its Jacobian
\begin{equation*}
  J(\epsilon)=\frac{\partial \vec{w}}{\partial \vec{y}}=\mathbb{I}+\epsilon \partial_y
  \vec{u}\, ,
\end{equation*}
 is non singular, namely that $\det J\neq 0$. This is certainly true if
 $\epsilon=0$. We could hence determine an upper bound for $\epsilon$ as:
 \begin{equation}
   \label{eq:epsilonstar}
   \epsilon^*=\min\{\epsilon>0: \forall q\in[0,\epsilon)\, ,\det J(q)\neq 0\}\, .
 \end{equation}

By using the form for $\vec{u}$ given by Eq.~\eqref{eq:uy} one can obtain
\begin{equation*}
  J(\epsilon)=\mathbb{I}_{2N}+\epsilon \partial_y \vec{u}=\mathbb{I}_{2N}+\epsilon\int_0^\tau\left[\mathcal{L}(r)-\langle\mathcal{L}\rangle\right]\,
  dr\, ,
\end{equation*}
and recalling the explicit form of $\mathcal{L}$ we get:
\begin{equation}
\label{eq:epsstartcond}
  \det (\mathbb{I}_{2N}+\epsilon \partial_y \vec{u})=0\Leftrightarrow \det
  \left(\mathbb{I}_{N}+\epsilon D_u \int_0^\tau\left[\mathbf{L}(r)-\langle\mathbf{L}\rangle\right]\,
  dr\right)\det
  \left(\mathbb{I}_{N}+\epsilon D_v \int_0^\tau\left[\mathbf{L}(r)-\langle\mathbf{L}\rangle\right]\,
  dr\right)=0\, .
\end{equation}

In the following we apply this strategy to compute $\epsilon^*$ for the twin network setting introduced in the main body of the paper.  

\subsection{The critical threshold for the twin network}
\label{sec:appl}

The analysis presented in the previous section~\ref{sec:epsilonstar} can be pushed further in the case of switching twin--networks (see main body of the paper for technical details on the model formulation). In this case we can in fact exaclty compute
$\langle\mathbf{L}\rangle=\gamma\mathbf{L}_1+(1-\gamma)\mathbf{L}_2$ and thus
the right hand side of Eq.~\eqref{eq:epsstartcond}. In fact let 
\begin{equation*}
  \phi(s)=\int_0^s\left[\mathbf{L}(r)-\langle\mathbf{L}\rangle\right]\, dr\, ,
\end{equation*}
then we have
\begin{equation*}
  \phi(s)=
  \begin{cases}
    (1-\gamma)(\mathbf{L}_1-\mathbf{L}_2)t &\text{if $\{t/T\}\in[0,\gamma)$}\\
    \gamma (T-t)(\mathbf{L}_1-\mathbf{L}_2)&\text{if $\{t/T\}\in[\gamma,1)$}
  \end{cases}\, ,
\end{equation*}
where $\{r\}$ denotes the fractional part of the real number $r$.

Let $\Lambda_{12}^\alpha$, $\alpha=1,\dots,N$, be the eigenvalues of $\mathbf{L}_1-\mathbf{L}_2$, then the roots of Eq.~\eqref{eq:epsstartcond} depend on $t$ and are of the form (we assume to order the eigenvalues such that $\Lambda_{12}^1=0$)
\begin{equation*}
\epsilon_\alpha=\frac{1}{-\Lambda_{12}^\alpha}\frac{1}{D_u\psi(t)}\text{ or }\epsilon_\alpha=\frac{1}{-\Lambda_{12}^\alpha}\frac{1}{D_v\psi(t)}\quad \forall \alpha=2,\dots,N\text{ and $t>0$}\, ,
\end{equation*}
where
\begin{equation*}
  \psi(t)=
  \begin{cases}
    (1-\gamma)t &\text{if $\{t/T\}\in[0,\gamma)$}\\
    \gamma(T-t)&\text{if $\{t/T\}\in[\gamma,1)$}
  \end{cases}\, ,
\end{equation*}
which has a minimum at $t=\gamma T$. Hence   Eq.~\eqref{eq:epsilonstar} returns:
\begin{equation}
\label{eq:epslow}
\epsilon^*= \frac{1}{\Lambda_{12}^N\gamma(1-\gamma)T}\min \left[\frac{1}{D_u},\frac{1}{D_v}\right]\, ,
\end{equation}
where we set  $\Lambda_{12}^N=\max_{\alpha} \lvert \Lambda_{12}^{\alpha}\rvert$, keeping in mind that the spectrum of $\mathbf{L}_1-\mathbf{L}_2$ contains both positive and negative eigenvalues.

\section{Another relevant application: the blinking network}
\label{sec:appl2}
Many interactions involving humans can be described by a binary contact lasting for a finite amount of time after which the two individuals get apart. During these events information can be mutually shared. If contacts are not mediated by electronic devices, viruses can pass from an individual to the other, hence triggering the epidemic spreading. As a first approximation, the dynamics of interaction within an large community can be broken down into a collection of successive pairwise exchanges, that extend over a finite window of time. Working in this setting, it is interesting to elaborate on the conditions (e.g. characteristic time of interactions) that discriminate between an homogenous or pattern like (in terms of virus load or information content)  asymptotic equilibrium. 

To this end we here consider a collection of networks, each made of $N$ isolated nodes with the exception of two nodes that are connected via an undirected link. Any two networks in this collection differ because a different link is activated. There are hence $N(N-1)/2$ different networks in total. All networks have the same Laplacian spectrum given by $\Lambda_1=0$ with multiplicity $N-1$ (i.e. the number of connected components the network is made of) and $\Lambda_N=-2$ with multiplicity $1$ (i.e. the minimal value the eigenvalues can assume in the case of complete bipartite network). 

Let us now define a periodically time varying network $\mathbf{A}(t)$ as follows. Fix a period $T>0$ and divide the interval $[0,T]$ into $m=N(N-1)/2$ equal disjoint subintervals $\tau_i$, $\cup_i \tau_i=[0,T]$ and $\tau_i\cap\tau_j=\emptyset$ if $i\neq j$. Hence, $|\tau_i|=\tau=T/m$ (of course this assumption can be relaxed and subintervals with different length allowed for). 

\begin{figure}[ht!]
\begin{center}
\includegraphics[width=0.85\textwidth]{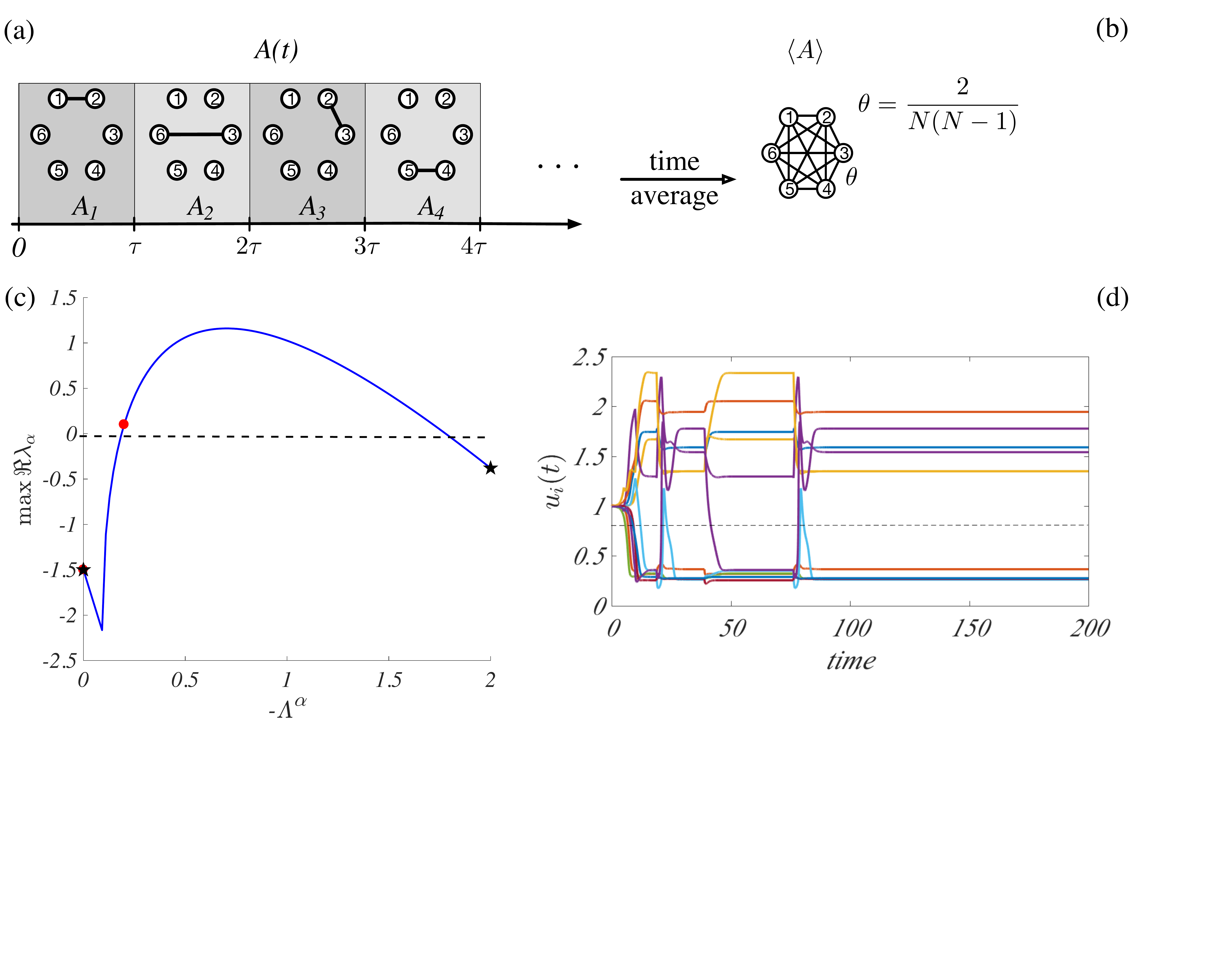}
\end{center}
\vspace{-3cm}
\caption{Blinking network. Panel a): $T$-periodic network built by using $m$ networks, $\mathbf{A}_i$ $i=1,\dots ,m$ ($m=N(N-1)/2$). In this example, each network is made by $N=6$ nodes among which only two are connected. Let $\tau=T/M$, then for $t\in[0,\tau)$ the $T$-periodic network coincides with $\mathbf{A}_1$, in the successive interval $t\in[\tau,2\tau)$, $\mathbf{A}(t)=\mathbf{A}_2$, and so on. The time varying network is then obtained by periodically repeating this construction. Panel b): the time average network $\langle \mathbf{A}\rangle =  K_N/M$, where $K_N$ denotes the complete network made by $N=6$ nodes, in this example. Panel c):  dispersion relation for the average network (red circles), for each static network $\mathbf{A}_i$ (black stars) ($N=11$ nodes) and for the continuous support case (blue curve). Panel d): the time evolution of the concentrations $u_i(t)$ for the blinking network composed by $N=11$ nodes. The reaction dynamics is given by the Brusselator model ($f(u,v)=1-(b+1)u+cu^2v$ and $g(u,v)=bu-cu^2v$ where $b$, $c$ are parameters of the model hereby fixed to $b=8$ and $c=10$); the homogeneous equilibrium is $\bar{u}=1$ and $\bar{v}=0.8$. The remaining parameters are $T=1$, $D_u = 3$ and $D_v = 10$.}
\label{fig:blinkA1}
\end{figure}

Assume to order the networks in the collection from $1$ to $N(N-1)/2$, then define $\mathbf{A}(t)$ by
\begin{equation}
\label{eq:Atblink}
\mathbf{A}(t)=\mathbf{A}_i \text{ if $t\in\tau_i$}\, ,
\end{equation}
where $\mathbf{A}_i$ is the adjacency matrix associated to one of the networks of the collection. Then we periodically repeat the procedure, i.e. if $t>T$ then consider $(t \mod T)$ instead of $t$. Differently stated, during each time interval, a novel link is created between a newly selected pair of nodes and the connection active during the preceding time window deleted. Hence, at each time there is one and only one active link (see top left panel of Fig.~\ref{fig:blinkA1} for an illustrative explanation). By making use of the aforementioned notation, we could also say $\mathbf{A}(t)=\mathbf{A}^{[\sigma(t)]}$, where $\sigma(t)=i$ if $t\in \tau_i$.
Because of the assumption of equal length intervals $\tau_i$, the adjacency matrix of the averaged network is given by
\begin{equation}
\label{eq:Atav}
\langle {A}\rangle_{ij}=
\begin{cases} \frac{2}{N(N-1)}\quad &\text{if $i\neq j$}\\
0\quad &\text{if $i=j$}
\end{cases}\, ,
\end{equation}
while the Laplacian matrix reads
\begin{equation}
\label{eq:Ltav}
\langle {L}\rangle_{ij}=
\begin{cases} \frac{2}{N(N-1)}\quad &\text{if $i\neq j$}\\
-\frac{2}{N}\quad &\text{if $i=j$}
\end{cases}\, ,
\end{equation}
whose spectrum is $\Lambda_1=0$ with multiplicity $1$ and $\Lambda_N=-2/(N-1)$ with multiplicity $N-1$. One can thus choose the network size $N$ to ensure that the relation dispersion admits a positive real part, once evaluated in $\Lambda_N=-2/(N-1)$ (see bottom left panel of Fig.~\ref{fig:blinkA1}).

To carry out one test we consider again the Brusselator model running on top of a such time varying network. The reaction terms read therefore $f(u,v)=1-(b+1)u+cu^2v$ and $g(u,v)=bu-cu^2v$, where $b$, $c$ are parameters of the model  fixed that we assign to be $b=8$ and $c=10$. The homogeneous equilibrium is thus  $u_*=1$ and $v_*=0.8$. The diffusion coefficients are set to $D_u = 3$ and $D_v = 10$.

We also assign the remaining model parameters in such a way that each network of the collection ($\mathbf{A}_i$) cannot yield Turing patterns. This amounts to requiring $\lambda_1 = \Re \lambda(\Lambda_1) < 0$ (the homogeneous equilibrium is stable) and $\lambda_N = \Re \lambda(\Lambda_N) < 0$ (see bottom left panel of Fig.~\ref{fig:blinkA1}).

Applying Theorem~\ref{thm:mainthm} one can find a positive $\epsilon^*$ such that for all $0<\epsilon<\epsilon^*$, the fast varying time dependent system~\eqref{eq:reac_diffacc3acc} exhibits Turing patterns. This is due to the fact that, under this operating condition, system~\eqref{eq:reac_diffacc3acc}  is close enough to its averaged analogue to be able to follow its orbits. Conversely,  system~\eqref{eq:reac_diffacc3acc} does not exhibit Turing patterns if $\epsilon$ is too large (see bottom right panel of Fig.~\ref{fig:blinkA1}).

\section{Non-periodic networks}
\label{sec:qpnet}

Our main result, Theorem~\ref{thm:mainthm}, holds true also if the time varying network is not periodic: as mentioned earlier, we solely require the existence of the 
generalised average
\begin{equation}
 \label{eq:ave2}
\langle\mathbf{L}\rangle=\lim_{T\rightarrow \infty}\frac{1}{T}\int_0^T \mathbf{L}(t)\, dt\, .
\end{equation}

The aim of this section is thus to shortly discuss some examples of a non-periodic time-varying networks for which Eq.~\eqref{eq:ave2} is well defined.

The first example exploits the non commensurability of the periods of two periodic networks. More precisely, let $G_1(t)$, respectively $G_2(t)$, be a $T_1$, respectively $T_2$, periodic network. Then, assuming $T_1/T_2\not\in\mathbb{Q}$, the network $G(t)=G_1(t)+G_2(t)$ is {\em quasi-periodic} and one easily obtains that
\begin{equation*}
\langle \mathbf{L}\rangle = \langle \mathbf{L}_1\rangle +\langle \mathbf{L}_2\rangle \, ,
\end{equation*}
where $ \langle \mathbf{L}_i\rangle $ is the average Laplacian matrix of the $i$--th network, computed over the period $T_i$.

Building on this setting one could generate interesting applications. Imagine that for all finite $t$, the network associated to $G(t)$ cannot develop patterns. Then we could speculate on the possibility that patterns arise 
in the averaged network in the limit $t\rightarrow \infty$. For instance, given two adjacency matrices $\mathbf{A}_1$ and $\mathbf{A}_2$ and two parameters $\gamma_j\in(0,1)$, $j=1,2$, we define
\begin{equation*}
G_j(t)=
\begin{cases}
\mathbf{A}_1 &\text{if $\{t/T_j\}\in[0,\gamma_j)$}\\
\mathbf{A}_2 &\text{if $\{t/T_j\}\in[\gamma_j,1)$}\, ,
\end{cases}
\end{equation*}
then $G(t)=G_1(t)+G_2(t)$ is quasi-periodic. For all finite $t$ no patterns are allowed on $G(t)$ if they are not observed on neither $\mathbf{A}_1$, $\mathbf{A}_2$ nor $\mathbf{A}_1+\mathbf{A}_2$. 
However, the average limit network $\langle G\rangle=(\gamma_1+\gamma_2)\mathbf{A}_1+(2-\gamma_1+\gamma_2)\mathbf{A}_2$ could still yield patterns for a suitable choice of $\gamma_j$. We do not provide an explicit numerical evidence of this mechanism here, but instead turn to consider a different model of non periodic networks. For this further example, a numerical realization will be also given to support the conclusion of the analysis.  

The final example that we are going to illustrate is based on aperiodic {\it words} (or binary strings) and follows the construction of a map associating to the {\it letters} (or entries) of the word a network. Let us consider for the sake of simplicity, two time periodic networks built on two different time windows of identical total duration:
\begin{equation}
P_0(t)=
\begin{cases}
\mathbf{A}_1 &\text{if $\{t/T\}\in[0,\gamma)$}\\
\mathbf{A}_2 &\text{if $\{t/T\}\in[\gamma,1)$}
\end{cases}
\text{ and }
P_1(t)=
\begin{cases}
\mathbf{A}_2 &\text{if $\{t/T\}\in[0,1-\gamma)$}\\
\mathbf{A}_1 &\text{if $\{t/T\}\in[1-\gamma,1)$}
\end{cases}
\end{equation}
where $T>0$ is the period and $\gamma\in(0,1)$ a fixed parameter.

Given a string made of $w=(i_{n})_{n\in\mathbb{N}}$, $i_n\in\{0,1\}$, we can finally obtain a time varying network by selecting $P_0$ or $P_1$ according to the entries appearing in $w$. The Laplacian matrix $L(t)$ associated to this latter network will be aperiodic if the word $w$ also is. On the other hand, because of the definition of $P_0$ and $P_1$ (they are based on windows of the same length) one straightforwardly obtains: $\left\langle \mathbf{L}\right\rangle =\gamma \mathbf{L}_{1}+(1-\gamma)\mathbf{L}_{2}$.

\begin{figure}[ht!]
\begin{center}
\includegraphics[width=0.65\textwidth]{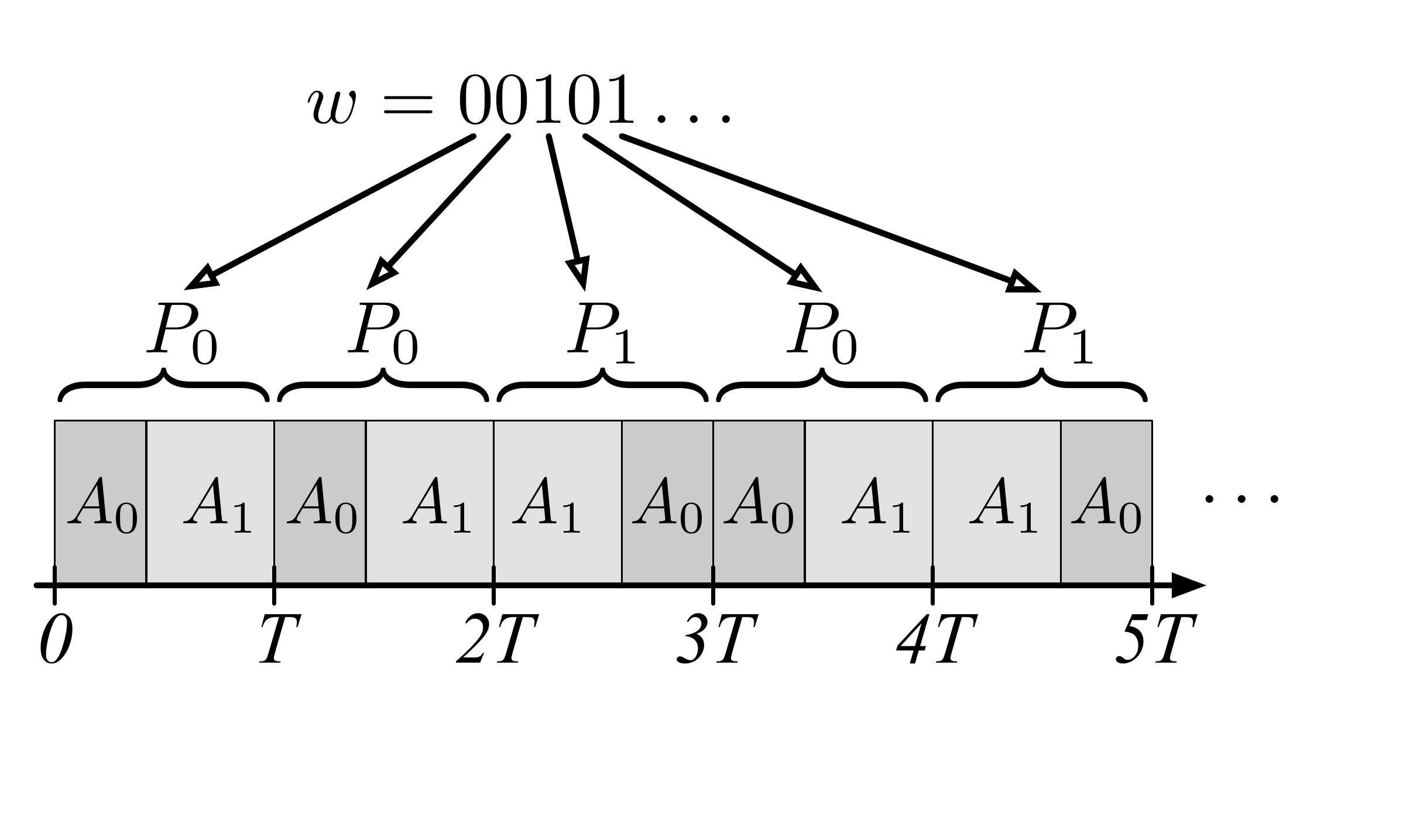}
\end{center}
\vspace{-2cm}
\caption{Aperiodic time varying network build using a Sturmian words. From the top, the word $w=00101\dots$, the map that associates to each letter (entry) of the word (binary string) the network $P_0(t)$ or $P_1(t)$, the resulting aperiodic time varying network.}
\label{fig:sturm}
\end{figure}

\begin{figure}[ht!]
\begin{center}
\includegraphics[width=0.85\textwidth]{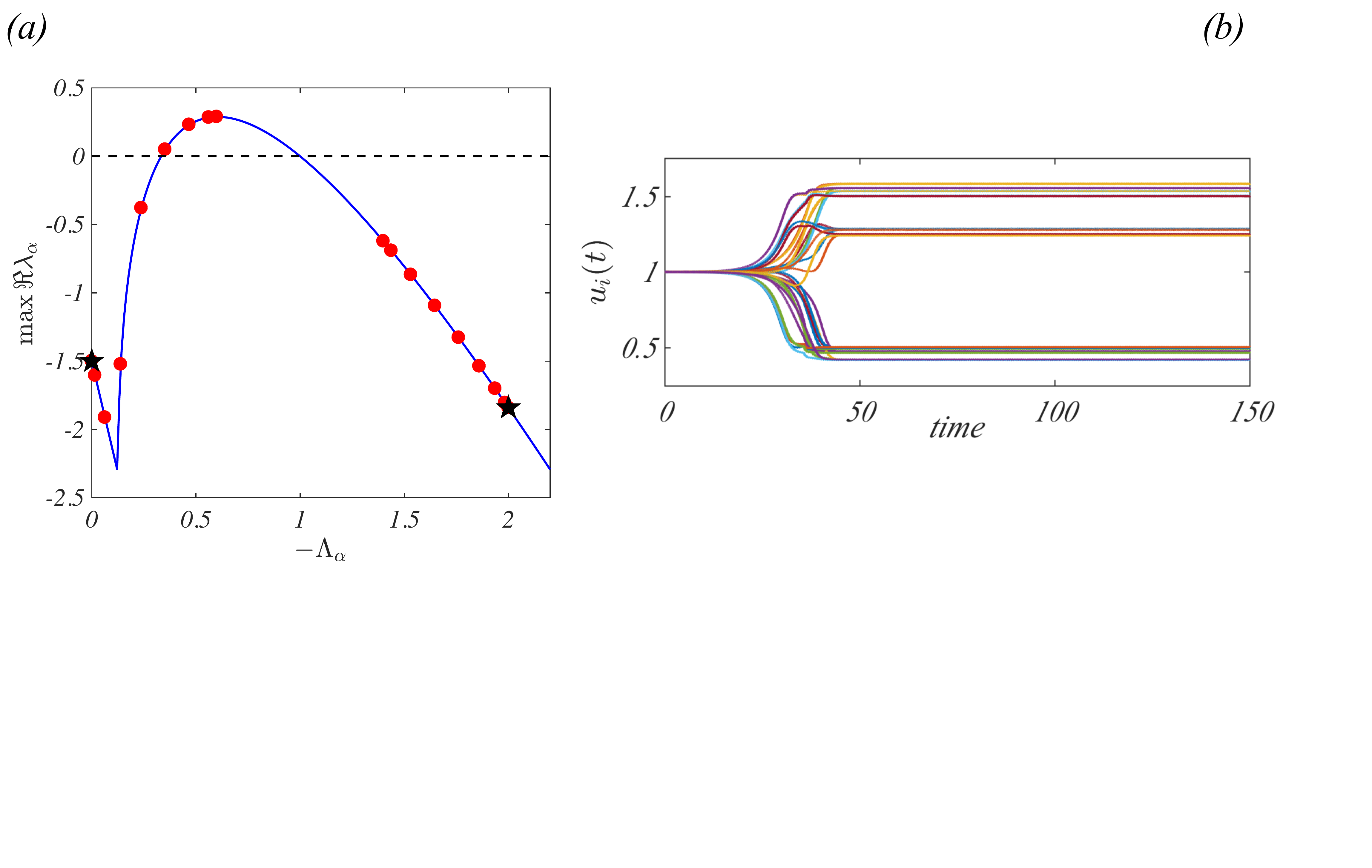}
\end{center}
\vspace{-3cm}
\caption{Twin network and Fibonacci words. Using the twin networks $\mathbf{A}_1$ and $\mathbf{A}_2$ (made by $N=32$ nodes) and the Fibonacci word $w^{\textit{Fibonacci}}=00101001001\dots$, we apply the scheme presented in Fig.~\ref{fig:sturm} to build a time varying aperiodic network $\mathbf{A}(t)$. Panel a): the dispersion relation ($\max\Re \lambda_{\alpha}$ vs. $-\Lambda(\alpha)$) for the average network (red circles), for each static twin network (black stars) and for the continuous support case (blue curve). Panel b): the time evolution of the concentrations $u_i(t)$ for the aperiodic network composed by $N=32$ nodes. The reaction dynamics is given by the Brusselator model ($f(u,v)=1-(b+1)u+cu^2v$ and $g(u,v)=bu-cu^2v$ where $b$, $c$ are parameters of the model hereby fixed to $b=8$ and $c=10$); the homogeneous equilibrium is $\bar{u}=1$ and $\bar{v}=0.8$. The remaining parameters are $T=1$, $D_u = 3$, $D_v = 10$ and $\epsilon=0.01$.}
\label{fig:fig5}
\end{figure}

Let us conclude by providing an explicit example (see also Fig.~\ref{fig:fig5}). Let us call the complexity of $w$ the application $p_{w}:\mathbb{N}\rightarrow\mathbb{N}$ which gives
for every $n$ the number of subwords of length $n$ in $w$. Using
the Morse-Hedlund theorem\footnote{A word $w$ is eventually periodic iff there exists $n\in\mathbb{N}$
such that $p_{w}(n)\leq n$. }, an aperiodic word $w$ of minimal complexity is such that 
\begin{equation}
\forall n\in\mathbb{N},\:p_{w}(n)=n+1.
\end{equation}
The words with minimal complexity are called Sturmian words~\footnote{Note that it is trivial to give examples of aperiodic words with higher complexity.}. A well-known example is the Fibonacci word $w^{\textit{Fibonacci}}=w_0w_1w_2\dots$ where the $n$--th subword, $w_n$, is given by 
\begin{equation}
w^{Fibonacci}(n)=2+\left\lfloor n\varphi\right\rfloor -\left\lfloor (n+1)\varphi\right\rfloor \quad\mbox{where}\quad\varphi=\frac{1+\sqrt{5}}{2}\, .
\end{equation}
Observe that such word can be obtained using the following recursion
\begin{align}
s_{0} & =0,\\
s_{1} & =01,\\
s_{n} & =s_{n-1}\cdot s_{n-2},\ \forall n\geq2,
\end{align}
where $s_{1}\cdot s_{2}$ is the concatenation of $s_{1}$ and $s_{2}$, from which the name Fibonacci follows. A numerical realization of the proposed scheme is reported in Figure \ref{fig:fig5}: patterns appear in the time varying network also if they are formally impeded on each subnetworks components.

%\subsection{An example based on weighted networks}
%
%Consider again the twin networks and their corresponding matrices
%$A_{1}$ and $A_{2}$. Define two periods $T_{1}$ and $T_{2}$ which
%are non commensurate, for example $T_{1}=1$ and $T_{2}=\sqrt{2}$.
%Let us define $L(t)$ by 
%\begin{equation}
%L(t)=\begin{cases}
%L_{1}+L_{3} & \mbox{if }\mod(t,T_{1})<\gamma\ \mbox{ and }\mod(t,T_{2})<\gamma\\
%L_{1}+L_{4} & \mbox{if }\mod(t,T_{1})<\gamma\ \mbox{ and }\mod(t,T_{2})\geq\gamma\\
%L_{2}+L_{3} & \mbox{if }\mod(t,T_{1})\geq\gamma\ \mbox{ and }\mod(t,T_{2})<\gamma\\
%L_{2}+L_{4} & \mbox{if }\mod(t,T_{1})\geq\gamma\ \mbox{ and }\mod(t,T_{2})\geq\gamma
%\end{cases}\label{eq:Ltweighted}
%\end{equation}
%where $L_{3}=\delta\cdot L_{1}$ and $L_{4}=\delta\cdot L_{2}$ with
%$\delta$ a small parameter. $L_{3}$ and $L_{4}$ can be viewed as
%perturbations of $L_{1}$ and $L_{2}$. We have
%\begin{equation}
%\left\langle L\right\rangle =(1+\delta)(\gamma L_{1}+(1-\gamma)L_{2}).%\end{equation}
%For $\delta$ small enough, the dispersion relation for every network
%configuration given by \eqref{eq:Ltweighted} is close to that corresponding
%to $L_{1}$ and $L_{2}$, and the dispersion relation of $\left\langle L\right\rangle $
%resembles that of $\gamma L_{1}+(1-\gamma)L_{2}$ as shown in figure
%\eqref{fig:fibo}, but the time-variation of $L(t)$ is aperiodic. 
%

\end{document}